\documentclass[reprint,aps,pre]{revtex4-2}
\usepackage{graphicx}
\usepackage{tikz}
\usepackage{mathtools}
\usepackage{amssymb}
\usepackage{textgreek}

\newcommand*\diff{\mathop{}\!\mathrm{d}}

\begin{document}
\title{Collapse transition of a Lennard-Jones polymer}

\author{Stefan Schnabel}

\author{Wolfhard Janke}

\affiliation{Institut f\"ur Theoretische Physik, Universit\"at Leipzig, IPF 231101, 04081 Leipzig, Germany}
\date{\today}




\begin{abstract}

Using the recently introduced parsimonious Metropolis algorithm bead-stick polymers both with infinite-range Lennard-Jones interaction and with truncation are simulated. The focus lays on determining the Boyle temperature for long chains with thousands of repeat units and on testing for theoretically predicted logarithmic corrections. Subsequently the behavior at the infinite-chain transition temperature, i.e., the $\Theta$-temperature is studied for chains with up to $N=32768$ repeat units by investigation of the scaling of the end-to-end distance, the radius of gyration, the specific heat, and their derivatives with $N$. 

\end{abstract}

\maketitle

\section{Introduction}
\label{sec-intro}

Flexible polymers in dilute solution undergo a coil-globule or $\Theta$-transition when the solvent is changed from good to bad or the temperature changes from high to low \cite{DeGennes,Cloizeaux}. The critical temperature is the infinite-chain limit of the Boyle temperature, the temperature at which the second virial coefficient vanishes \cite{flory1953principles}. This transition is expected to be of second order with the ``size'' of the polymer in the form of the end-to-end distance changing continuously. At the critical point the polymer behaves on large enough scales like an ideal chain or Gaussian random walk. At higher and lower temperature the chain resembles a self-avoiding walk and a collapsed globule, respectively. Being a tricritical point \cite{DeGennes_JPL}, the upper critical dimension of the $\Theta$-transition is $d_c=3$ \cite{DeGennes,Cloizeaux}, logarithmic corrections to scaling should be present and have been calculated to first order \cite{Duplantier_JPF,Duplantier_EPL,Duplantier_JCP} and later to second order \cite{Hager} by means of renormalization-group methods. 

The predicted logarithmic corrections have not been verified by means of Monte Carlo (MC) simulations which is due to the fact that the logarithmic asymptotic scaling manifests clearly only for very large systems which are difficult to handle. In the past chains with thousands of repeat units could only be investigated in the form of self-attracting lattices walks \cite{GrassbergerHegger,Grassberger} and off-lattice systems with only some dozens \cite{Vassilis} or few hundreds \cite{Yong,Rubio1,Rubio2} of monomers have been realized. Due to these limitations studies determining $\Theta$-temperatures are few and far between even though there is great interest in transitions of single polymer chains and their phase diagram. For more recent works see for instance \cite{Rampf,Zhang}. In an attempt to address the problem we recently introduced a MC method \cite{poly_tree} that substantially accelerates the simulation of long polymer chains with continuous degrees of freedom and extends the length of chains that can be treated by at least an order of magnitude. In this study we now employ this method to investigate the $\Theta$-transition of a Lennard-Jones polymer.

The paper is organized as follows: In Sec. \ref{sec-model} we review the model and geometric observables. In Sec. \ref{sec-methods} we discuss the MC methods we use, how the second virial coefficient is measured, and how averages of observables based on energy can be measured very efficiently. The outcome of our simulations is then presented and discussed in Sec. \ref{sec-res}. In Sec. \ref{sec-conc} we finish with a summary and some concluding remarks.


\section{Model}
\label{sec-model}

We consider a bead-stick chain $\mathbf{X}=(\mathbf{x}_1,\dots,\mathbf{x}_N)$ in three dimensions where the monomers $\mathbf{x}_i$ interact pairwise via a 12-6 Lennard-Jones (LJ) interaction:
\begin{equation}
U(r)=4\epsilon\left[ \left(\frac\sigma r\right)^{12} - \left(\frac\sigma r\right)^{6} \right].
\label{eq:LJ}
\end{equation}

The chain is fully flexible, but the bonds connecting adjacent monomers are of fixed length:
\begin{equation}
|\mathbf{x}_k-\mathbf{x}_{k-1}| = b.
\end{equation}
The monomer diameter is adjusted to $\sigma=2^{-1/6}b$ such that $U_{\rm LJ}(b)=-\epsilon$.

The total energy function for an isolated chain reads

\begin{equation}
E=\sum_{i=1}^{N-1}\sum_{j=i+1}^{N}U(|\mathbf{x}_i-\mathbf{x}_j|).
\label{eq:chain_energ}
\end{equation}

\begin{figure}[t]
\begin{center}
\includegraphics[width=0.95\columnwidth]{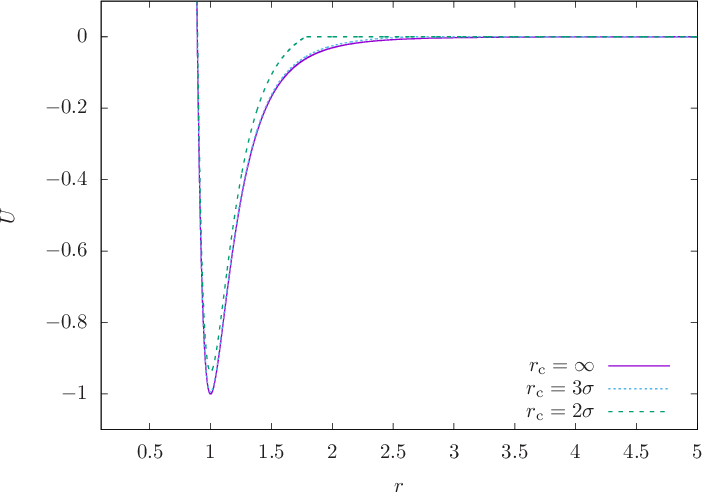}
\caption{\small{\label{fig:traj_comp} LJ potentials (\ref{eq:LJ},\ref{eq:LJc}) without truncation and with truncation at $r=r_{\rm c}$.}}
\end{center}
\end{figure}

As a variation of the model we will also use the truncated and shifted LJ potential
\begin{equation}
U_{\rm c}(r)=U\left(\min(r,r_{\rm c})\right)-U(r_{\rm c}).
\label{eq:LJc}
\end{equation}
where $r_{\rm c}$ is the cut-off distance. The shapes of the resulting potentials are depicted in Fig.\ \ref{fig:traj_comp}.

Two geometric observables are commonly considered: The end-to-end distance $r_{\rm ee}=|\mathbf{x}_N-\mathbf{x}_1|$ and the radius of gyration
\begin{equation}
r_{\rm gyr}^2=\frac1N\sum\limits_{i=1}^N\left(\mathbf{x}_i - \frac1N\sum\limits_{j=1}^N\mathbf{x}_j\right)^2.
\end{equation}

In the following we will present distances in units of $b$ and temperatures in units of $k_{\rm B}/\epsilon$.

\section{Methods}
\label{sec-methods}

\subsection{Conformational updates}

To generate new configurations we apply three different MC moves.
First the pivot move \cite{pivot} is used. It randomly picks a monomer as a fulcrum and rotates all monomers on one side, i.e., 
\begin{equation}
\mathbf{x}'_i=\mathcal{R}(\mathbf{x}_i-\mathbf{x}_k)+\mathbf{x}_k \quad \text{for all} \quad i>k
\end{equation}
where $\mathcal{R}$ is a random rotation matrix.
Second we use a bond rotation, where a single bond $\mathbf{b}_k=\mathbf{x}_{k+1}-\mathbf{x}_k$ is randomly assigned a new direction while all other bonds remain the same. Consequently
\begin{equation}
\mathbf{x}'_i=\mathbf{x}_i+\mathbf{b}'_k-\mathbf{b}_k \quad \text{for all} \quad i>k.
\end{equation}
If only a single isolated chain is considered the system is symmetric under rotation and translation which allows to either perform both moves as prescribed above or to apply the inverse transformations to the other part of the polymer with monomer indices $i\le k$. 

As we will discuss below, in order to determine the second virial coefficient, it is necessary to simulate two interacting chains. In this situation we additionally employ a generalized crank-shaft move: We select two monomers at positions $\mathbf{x}_k$ ad $\mathbf{x}_l$, determine the axis through both of them, and rotate all monomers in between:
\begin{equation}
\mathbf{x}'_i=\mathcal{R}(\mathbf{x}_i-\mathbf{x}_k)+\mathbf{x}_k \quad \text{for all} \quad k<i<l
\end{equation}
where $\mathcal{R}$ is a random rotation with the constraint $\mathbf{x}_k-\mathbf{x}_l=\mathcal{R}(\mathbf{x}_k-\mathbf{x}_l)$. We bias the selection of monomers such that predominantly small parts of the polymer are modified.

Since this study focuses on the collapse transition, i.e., on the two high-temperature phases, the more elaborate moves designed to sample the crystal-like low-temperature state \cite{poly_method} are not needed.

\subsection{Binary tree data-structure}

Clisby \cite{Clisby1} introduced a binary tree data-structure for the efficient sampling of self-avoiding walks on lattices which allowed the determination of scaling exponents with very high accuracy \cite{Clisby1,Clisby2,Clisby3}. Recently, the method has been generalized for the simulation of off-lattice polymers \cite{poly_tree}. In the data tree it employs each node relates to a piece of the chain. The root node represents the entire polymer, its children are the polymer's two halves, each child node of the root's children stands for a quarter and so on. With each level down the tree the number of nodes doubles while the amount of monomers in each node is reduced by a factor of two. The nodes do not store the actual monomer positions but rather the information of a simple geometrical body that contains all the node's monomers. While on a cubic lattice a natural choice for this body is a cuboid, in the present case of continuous degrees of freedom we use spheres. For the tree's leaves, i.e., for individual monomers these spheres are situated at the monomer position $\mathbf{x}_k$ and they have radius zero. At higher levels spheres are constructed recursively by finding the smallest sphere that contains the spheres of the child nodes.

If a node $A$ contains $n_A$ monomers within a sphere of radius $r_A$ around the center point $\mathbf{c}_A$ then the interaction $E_{AB}$ between two nodes $A$ and $B$ with disjoint sets of monomers can be estimated. If the two spheres are distant this can be done with good precision and little computational effort and as we will discuss further below an efficient MC method can be based on this idea. Defining the extremal distances
\begin{eqnarray}
d^{\rm min}_{AB}=|\mathbf{c}_A-\mathbf{c}_B|-r_A-r_B,   \\
d^{\rm max}_{AB}=|\mathbf{c}_A-\mathbf{c}_B|+r_A+r_B
\end{eqnarray}
we find that if $d^{\rm min}_{AB}\ge2^{1/6}\sigma$
\begin{equation}
E_{AB} \in \left[ n_A n_B U(d^{\rm min}_{AB}),\, n_A n_B U(d^{\rm max}_{AB}) \right],
\end{equation}
since under this condition the potentials (\ref{eq:LJ}) and (\ref{eq:LJc}) are monotonically increasing throughout $[d^{\rm min}_{AB},d^{\rm max}_{AB}]$. If the uncertainty is too large, i.e., if the interval of possible values for  $E_{AB}$ is too wide and narrower boundaries are desired, it is possible to ``increase the resolution'' by descending one branch of the tree. If we label the child nodes of $A$ as $A_{\rm l}$ and $A_{\rm r}$ we obtain
\begin{eqnarray}
E_{AB} \in \left[ n_{A_{\rm l}} n_B U(d^{\rm min}_{{A_{\rm l}}B})+n_{A_{\rm r}} n_B U(d^{\rm min}_{{A_{\rm r}}B}),\right. \quad \nonumber\\
     \quad \left. n_{A_{\rm l}} n_B U(d^{\rm max}_{{A_{\rm l}}B})+n_{A_{\rm r}} n_B U(d^{\rm max}_{{A_{\rm r}}B}) \right]
\label{eq:node_split}
\end{eqnarray}
if $d^{\rm min}_{{A_{\rm l}}B}\ge2^{1/6}\sigma$ and $d^{\rm min}_{{A_{\rm r}}B}\ge2^{1/6}\sigma$. Of course, instead of $A$, node $B$ can be replaced by its children as well. Intuitive heuristics are to split the node that contains the greater number of monomers or the node with the greater sphere, both of which are sound strategies leading to similar performance.

The internal energy of any node $A$ is recursively given by
\begin{equation}
E_A=E_{{A_{\rm l}}{A_{\rm r}}}+E_{A_{\rm l}}+E_{A_{\rm r}},
\label{eq:int_energ}
\end{equation}
and the total energy (\ref{eq:chain_energ}) is the internal energy of the tree's root node. In virtually all cases the straightforward recursive evaluation of eq.~(\ref{eq:int_energ}) will not directly lead to an estimate of $E_A$ since many adjacent nodes will have overlapping spheres or minimal distances smaller than $2^{1/6}\sigma$. Hence, applications of the refinement described in eq.~(\ref{eq:node_split}) are required. However, even when all involved node-node interactions can be estimated initially the uncertainty of the resulting sum will often be too large such that further improvement is necessary. We apply two techniques to control the precision.

First one can choose to accept only estimations of node-node interactions that do not exceed a given uncertainty $\kappa$ per monomer pair, i.e., both $d^{\rm min}_{CD}\ge2^{1/6}\sigma$ and $U(d^{\rm max}_{CD}) - U(d^{\rm min}_{CD}) \le \kappa $ must be true or the interaction between nodes $C$ and $D$ is further decomposed by splitting either $C$ or $D$. This has the advantage that decisions can be made locally without reference to the global picture such that a recursive evaluation of eq.~(\ref{eq:int_energ}) with minimal memory requirement is possible. However, if done this way it is not possible to easily improve the precision of the sum once the calculation is completed. The entire process has to be run again with a smaller $\kappa$.

Alternatively, one can keep track of all node-node interactions that currently contribute to the sum and sort them according to their respective uncertainties. Thereby, it is not necessary to establish strict order, instead it suffices to group the interactions such that the uncertainties of interactions in each group are roughly similar, e.g., do not differ by more than a factor of two. It is then possible to repeatedly select an interaction $E_{CD}$ with a large uncertainty and replace it by the two node-node interactions obtained when either $C$ or $D$ is split thus improving the global estimate until a satisfactory precision is reached.

The binary tree also proves very useful in the case of the truncated potential. Then the interaction between two nodes $A$ and $B$ can be recursively evaluated exactly:
\begin{eqnarray}
E_{AB} &=& E_{{A_{\rm l}}B}+E_{{A_{\rm r}}B} \qquad \text{or}\\
E_{AB} &=& E_{A{B_{\rm l}}}+E_{A{B_{\rm r}}},
\end{eqnarray}
with the recursion ending if either $d_{CD}^{\rm min} \ge r_{\rm c}$ meaning that there is no interaction between the two nodes $C$ and $D$ or if $n_C=n_D=1$ in which case the monomer-monomer potential is directly calculated. This procedure can be seen as an efficient way to determine which of the $n_An_B$ monomer pairs are closer than the cutoff radius $r_{\rm c}$ and contribute to $E_{AB}$.

\subsection{The parsimonious Metropolis algorithm}

When the canonical ensemble is sampled using the Metropolis algorithm \cite{Metropolis} the acceptance of a proposed update from state $\mu$ to state $\nu$ is given by
\begin{equation}
P^{\rm acc}_{\mu\rightarrow\nu}=\min(1,P_\nu/P_\mu)=\min(1,e^{-\beta\Delta E}),
\end{equation}
where $\Delta E =E_\nu-E_\mu$ is the difference in energy between the two states and $\beta=(k_{\rm B}T)^{-1}\ge0   $ the inverse temperature.
The standard method to decide whether an update is to be accepted usually involves first checking if $\Delta E \le 0$ which directly implies acceptance.
If on the other hand $\Delta E > 0$ one finds $P^{\rm acc}_{\mu\rightarrow\nu}\in (0,1)$ and a decision is made by drawing a random number $\xi\in[0,1)$ and accepting the update if $\xi<P^{\rm acc}_{\mu\rightarrow\nu}$.
This procedure is especially suitable if the change in energy can be calculated very fast while drawing a random number is in comparison computationally expensive. A typical example for such a combination are spin flips in the (short-range) Ising model and the widely used Mersenne Twister random number generator.
If on the other hand the calculation of $\Delta E$ is the more demanding part as for instance in the case of long-range interacting systems it can be beneficial to use an alternative technique:
Draw the random number $\xi\in[0,1)$ first and determine $\eta=-\beta^{-1}\ln\xi$.
Since an update is accepted if $\Delta E<\eta$ and rejected otherwise, the task at hand is no longer to calculate the change in energy with maximal precision, but rather to establish whether this inequality is true or false. As we have demonstrated above, strict boundaries for the energy can be established and refined very efficiently using the binary tree data-structure. Further details of the algorithm can be found in \cite{poly_tree} and \cite{Fabio}.

\subsection{Measuring energetic observables}

As we have indicated, the technique we are applying for the simulation of LJ polymers with untruncated potential does not require the exact calculation of the changes in energy associated with each MC step and it is, therefore, very fast. However, we are, of course, interested in quantities like the average energy $\langle E \rangle$, the second moment $\langle E^2 \rangle$, and averages of products with other observables, e.g. $\langle r_{\rm ee} E \rangle$ or more general averages of observables of the form $\langle O(E,\dots) \rangle$. In order to minimize the statistical error it is desirable to measure these observables as often as possible, yet the huge computational effort ($\propto N^2$) required to calculate $E=\sum_{i=1}^{N-1}\sum_{j=i+1}^{N}U(|\mathbf{x}_i-\mathbf{x}_j|)$ does not allow this. However, we can estimate $E$ very efficiently as long as the precision is low, i.e., with much less effort we can measure an energy-like observable $\mathcal{E}_\delta$ that fulfills the condition $|\mathcal{E}_\delta-E|<\delta$.

If $\delta$ can be chosen freely we can express the average energy for instance in the following way:
\begin{equation}
\langle E \rangle = \langle \mathcal{E}_{10^{-1}} \rangle + \langle \mathcal{E}_{10^{-4}} - \mathcal{E}_{10^{-1}} \rangle + \langle E - \mathcal{E}_{10^{-4}} \rangle.
\end{equation}
The first term on the right can be seen as an imprecise estimate, the second is its correction, and the third a correction of this correction. While measuring these averages our goal should be to end up with comparable statistical uncertainties for all of them. While $\mathcal{E}_{10^{-1}}$ has a wide distribution the many measurements that are needed for a precise determination of $\langle \mathcal{E}_{10^{-1}} \rangle$ do not slow down the simulation since the large $\delta=10^{-1}$ allows for a fast evaluation. Next, the distribution of $\mathcal{E}_{10^{-4}} - \mathcal{E}_{10^{-1}}\in(-10^{-1}-10^{-4},10^{-1}+10^{-4})$ is much more narrow. Consequently, we need fewer measurements each of which, however, takes a longer time. Finally although calculating $E - \mathcal{E}_{10^{-4}}$ is very demanding only very few measurements are needed since its distribution is very narrow.  In the present example its width of order $10^{-4}$ or less might even be smaller than the statistical uncertainty of $\langle \mathcal{E}_{10^{-1}} \rangle$ such that the systematic impact on the uncertainty of $\langle E \rangle$ would be negligible.

In the same way we can calculate averages $\langle O(\mathcal{E}_{\delta_1},\dots) \rangle$, $\langle O(\mathcal{E}_{\delta_2},\dots)-O(\mathcal{E}_{\delta_1},\dots) \rangle$, etc., although suitable values for $\delta_1,\delta_2,\dots$ might be different for different observables. This method is a natural extension of our simulation method and several bits of previously developed computer code could be reused for its implementation.

\subsection{Measuring the second virial coefficient}

At the Boyle temperature $T_{\rm B}$ the second virial coefficient $A_2$ vanishes
\begin{equation}
A_2(T_{\rm B}(N),N) = 0
\end{equation}
and the $\Theta$-point with predicted tricritical behavior is the Boyle temperature of the infinite chain,
\begin{equation}
\Theta=\lim_{N\rightarrow\infty}T_{\rm B}(N).
\end{equation}
Therefore, the Boyle temperature is to be understood as the analog of the Boyle point of a real gas, i.e., the temperature where the gas is describrd by the ideal gas relation $p/nRT=1/V$, except for higher-order corrections (third order and higher) in $1/V$ \cite{flory1953principles}. To determine the temperature $\Theta$ precise measurements of the second virial coefficient $A_2$ are essential. The latter can be expressed as
\begin{equation}
A_2(T,N)=\frac{\int e^{-\frac{E(\mathbf{X})+E(\mathbf{Y})}{k_{\rm B}T} }\left(1-e^{-\frac{E_{\rm inter}(\mathbf{X},\mathbf{Y})}{k_{\rm B}T}}\right)\diff\mathbf{X} \diff\mathbf{Y}}{2Z(N,T)^2},\\
\end{equation}
where $E_{\rm inter}$ is the interaction between two polymer chains,
\begin{equation}
    E_{\rm inter}(\mathbf{X},\mathbf{Y})=\sum_{i=1}^{N}\sum_{j=1}^{N}U(|\mathbf{x}_i-\mathbf{y}_j|),
\end{equation}
and $Z$ the partition function of an isolated chain
\begin{equation}
    Z(N,T)= \int e^{-\frac{E(\mathbf{X})}{k_{\rm B}T} }\diff\mathbf{X}.
\end{equation}
Using the distance between the two center of masses
\begin{equation}
    r_{\rm cm}(\mathbf{X},\mathbf{Y})=\frac1N\left|\sum\limits_{i=1}^N(\mathbf{x}_i-\mathbf{y}_i)\right|,
\end{equation}
one can also write
\begin{equation}
A_2(T,N)=2\pi\int_0^\infty r_{\rm cm}^2\left\langle1-e^{-\frac{E_{\rm inter}}{k_{\rm B}T}}\right\rangle_{r_{\rm cm}} \diff r_{\rm cm}, \label{eq:E_II_av}
\end{equation}
where the average is taken for fixed values of $r_{\rm cm}$. 

The most commonly applied method for measuring $A_2$ simulates two independent chains $\mathbf{X}$ and $\mathbf{Y}$ at temperature $T$ in order to generate independent pairs of configurations $(\mathbf{X}_t,\mathbf{Y}_t)$. For each pair $E_{\rm inter}$ is sampled by placing the two chains at different distances and orientations with respect to each other and measuring the interaction. This can be done specifically by first shifting the two polymers such that their centers of gravity reside at the origin. Then a random rotation is applied to one of them and finally one chain is shifted to a random point within a sphere around the origin. The radius of the sphere has to be large enough to include all distances that allow relevant interaction between the two polymers. It is easy to see that the more frequent occurrence of large distances renders this simple method somewhat ineffective and an obvious improvement would be the introduction of a weight function and consequently a weighted average to enhance the selection of smaller distances. Regardless of this issue, the main problem with this strategy is that there are rare cases where the two chains interact strongly  leading to a comparatively large negative $E_{\rm inter}$. Since $-E_{\rm inter}$ appears in the exponent in (\ref{eq:E_II_av}) these cases have an extremely high impact, but for longer chains they are practically impossible to find with the described method since as far as the measurement of $E_{\rm inter}$ is concerned it performs only simple sampling. For this reason we only apply this method for polymers with $N\le32$.

A method for investigating longer chains in the case of interacting self-avoiding lattice walks has been provided by Grassberger and Hegger \cite{GrassbergerHegger} in the form of importance sampling. They propose to simulate an ensemble where the two chains interact and microstates occur with a probability proportional to the weight function

\begin{equation}
W_{\rm GH}(\mathbf{X},\mathbf{Y}) = e^{-\frac{E(\mathbf{X})}{k_{\rm B}T}}e^{-\frac{E(\mathbf{Y})}{k_{\rm B}T}}e^{-\frac{E_{\rm inter}(\mathbf{X},\mathbf{Y})}{k_{\rm B}T}}.
\end{equation}
Now the states with negative $E_{\rm inter}$ are sampled much more frequently and through reweighting they contribute to the average (\ref{eq:E_II_av}) with equal impact. We choose to modify this weight function for the model considered here. The contributions from the individual chains remain
\begin{equation}
W(\mathbf{X},\mathbf{Y})=e^{-\frac{E(\mathbf{X})}{k_{\rm B}T}}e^{-\frac{E(\mathbf{Y})}{k_{\rm B}T}}W_{\rm inter}(\mathbf{X},\mathbf{Y})
\end{equation}
and for the contribution of the inter-chain interaction we use in case of the truncated potential
\begin{equation}
W_{\rm inter}(\mathbf{X},\mathbf{Y})=
        \begin{cases}
            e^{-\frac{E_{\rm inter}(\mathbf{X},\mathbf{Y})}{k_{\rm B}T}} \, \text{if} \,\, E_{\rm inter}<-10^{-6}\epsilon, \\
            1.0 \qquad \qquad\, \text{if} \,\, E_{\rm inter}>10^{-6}\epsilon, \\
            e^{-r_{\rm cm}(\mathbf{X},\mathbf{Y})/\rho} \quad \text{otherwise},
        \end{cases}
\end{equation}
where $r_{\rm cm}$ is again the distance between the two center of masses. Now the regions of state space with positive chain-chain interaction -- which do not exist for Grassberger and Hegger's model -- are not suppressed and do not form barriers hampering the simulation. To prevent the two chains from permanently losing contact, the distance between the two chains is now constrained by the term $\exp[-r_{\rm cm}(\mathbf{X},\mathbf{Y})/\rho]$ in the weight function, if the inter-chain interaction is close to zero. Here $\rho\propto\sqrt{N}$ is a scale factor.

The case of non-truncated potential is more complicated. As with the energy of the individual chains, it is highly desirable to avoid calculating or updating the full chain-chain interaction with each MC move due to the high computational cost. Therefore, for generating the weight $W_{\rm inter}$ we opted to use $\alpha E_{\rm inter}^{c}$ as a proxy for $E_{\rm inter}$. Here, $E_{\rm inter}^{c}$ is the inter-chain interaction with a cutoff radius of $2b\approx2.2\sigma$ and $\alpha$ is a factor adjusted such that at the temperatures simulated $\max\{\exp[-(E_{\rm inter}-\alpha E_{\rm inter}^{c})/k_{\rm B}T]\}\approx 1$. The actual measurements are of course taken for the correct quantity $E_{\rm inter}$, are appropriately reweighted, and are accelerated by the means described in the previous subsection.


\section{Results and Discussion}
\label{sec-res}

\subsection{Logarithmic corrections and $N_0$}

In his theoretical analysis of the $\Theta$-transition, Duplantier \cite{Duplantier_JPF,Duplantier_EPL,Duplantier_JCP} modeled the polymer as a continuous path of length $S$ and introduced a ultra-violet cutoff $s_0$ marking the minimal distance along the path for which the polymer can interact with itself. Then, through $S/s_0\rightarrow\infty$ the thermodynamic limit is approached. As a consequence scaling relations are expressed as functions of $S/s_0$ and while $s_0$ can be absorbed by the prefactor of algebraic terms $a(S/s_0)^x=a'S^x$, its appearance in terms of logarithmic corrections cannot be avoided.

Universality dictates that asymptotically the results of Duplantier must also be correct for the model we consider here, if we replace $S/s_0$ by $N/N_0$ which raises the questions of the value of the constant $N_0$. Naively, the fact that interacting monomers must be separated by at least two bonds suggests $N_0=2$. However, in an imagined renormalization process one would expect this value to be subject to change. Unfortunately, the data we obtain is not precise enough and the chains we can investigate are not long enough to reliably determine $N_0$ through fits. One would expect $N_0$ to remain of the order of one and this seems to agree with our results.

\subsection{Determination of the transition temperature $\Theta$}

We performed simulations for four different cutoff radii ($r_{\rm c}/\sigma\in\{2, 3, 4, 5\}$) and the untruncated potential ($r_{\rm c}=\infty$). We investigated chain lengths $N\in\{2, 3,\dots,32\}$ using the simple-sampling method and $N\in\{8\times 2^k, 12\times 2^k, 19\times 2^k\}$ with integer $k\ge0$ using importance sampling. For $r_{\rm c}=2\sigma$ it is $N\le8192$ and for the other cases $N\le4096$.

\begin{figure}
\begin{center}
\includegraphics[width=0.95\columnwidth]{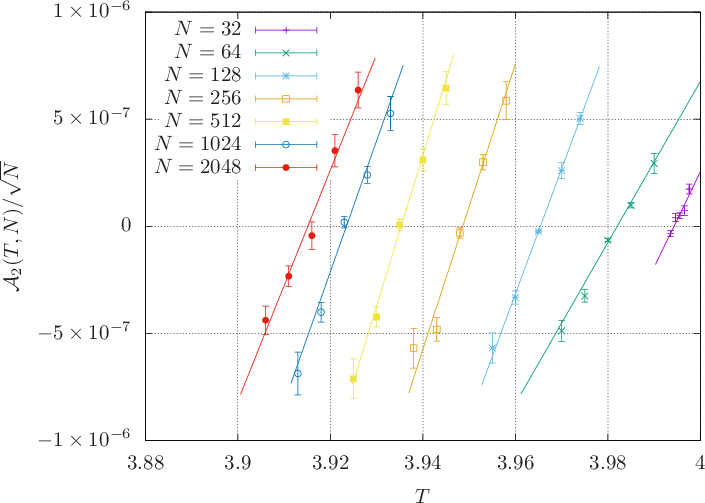}
\caption{\small{\label{fig:virial2} Some measured values for $\mathcal{A}_2(T,N)$ and linear fits for the untruncated potential ($r_{\rm c}=\infty$).}}
\end{center}
\end{figure}

For the determination of the Boyle temperature for each system we selected five temperatures in a relatively small interval that includes $T_{\rm B}$ and measure
\begin{equation}
    \mathcal{A}_2(T,N) \coloneqq \left\langle\frac{1-e^{-\frac{E_{\rm inter}(\mathbf{X},\mathbf{Y})}{k_{\rm B}T}}}{W_{\rm inter}(\mathbf{X},\mathbf{Y})}\right\rangle
\end{equation}
in the ensemble described above. Technically, due to a difference in normalization $\mathcal{A}_2(T,N)\ne A_2(T,N)$ but $\mathcal{A}_2(T,N)=0$ does imply $A_2(T,N)=0$. As can be seen in Fig.~\ref{fig:virial2} using small temperature intervals has two advantages. On the one hand the dependency on $T$ is approximately linear and $T_{\rm B}$ can be determined without reference to theoretical predictions while  on the other hand it appears that for simulations of equal length the statistical uncertainty of $\mathcal{A}_2$ becomes smallest in the proximity of $T_{\rm B}$ where $\mathcal{A}_2\approx0$. We obtained estimates of the statistical errors of $T_{\rm B}$ via the fitting process and due to the relatively small number of data points they themselves are relatively uncertain.

\begin{figure}
\begin{center}
\includegraphics[width=0.95\columnwidth]{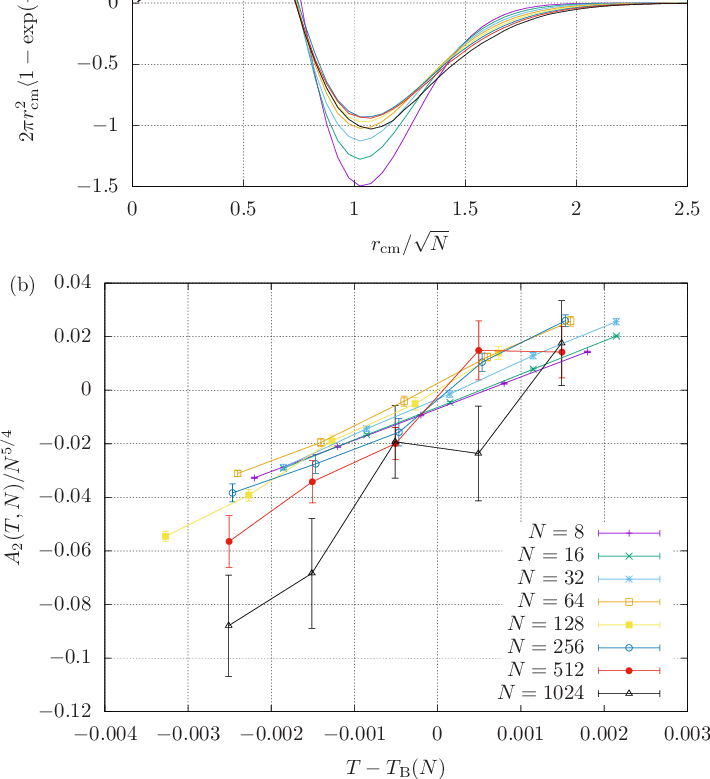}
\caption{\small{\label{fig:virial2_true} (a) Rescaled contributions to the second virial coefficient for different sizes $N$ at $T_{\rm B}(N)$. (b) The second virial coefficient $A_2$ in the proximity of the Boyle temperature $T_{\rm B}(N)$. (The slight deviation from $A_2(T_{\rm B}(N),N)=0$ is caused by systematic binning errors.) For both plots $r_{\rm c}=2\sigma$. Both of the exponents of $N$ in the ordinates have been chosen merely for convenience and do not signify suspected asymptotic behavior.}}
\end{center}
\end{figure}

For systems with truncated potential where $E_{\rm inter}$ is measured exactly every time we also determined the distance-dependent averages used in eq.\ (\ref{eq:E_II_av}) which are shown in Fig.~\ref{fig:virial2_true}a. In principle these allow the calculation of the true second virial coefficient $A_2(T,N)$ (Fig.~\ref{fig:virial2_true}b). However, even with truncated potential we used $\mathcal{A}_2(T,N)$ to determine $T_{\rm B}$ since the binning process needed to obtain $A_2(T,N)$ this way can introduce systematic errors. 

\begin{figure}[t]
\begin{center}
\includegraphics[width=0.95\columnwidth]{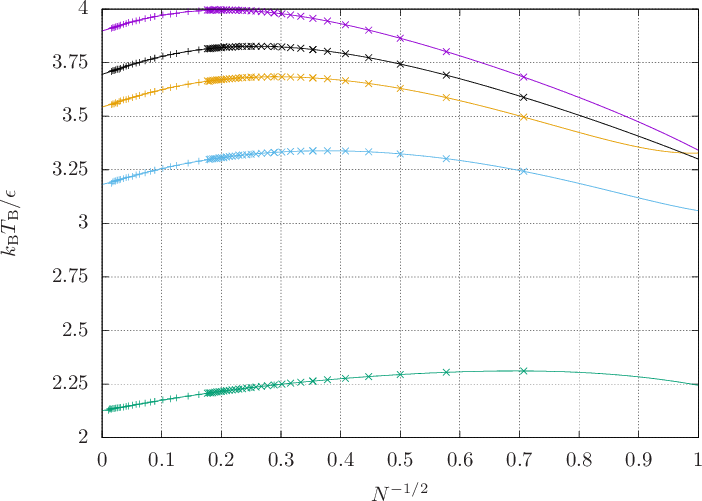}
\caption{\small{\label{fig:boyle} Boyle temperatures for chains of different lengths $N$. From top to bottom the cutoff radius $r_{\rm c}$ equals $\infty,5\sigma,4\sigma,3\sigma$, and $2\sigma$.} Values are obtained by simple sampling ($\times$) and importance sampling ($+$).
Lines show the fitted fourth order polynomials with logarithmic corrections $q_l(N)$.
}
\end{center}
\end{figure}

The Boyle temperatures we obtained through linear fits of $\mathcal{A}_2$ are displayed in Fig.~\ref{fig:boyle}. As expected the introduction and reduction of the cutoff radius reduces the transition temperature substantially. Even a relatively distant cutoff at $r_{\rm c}=5\sigma$ where $U(r_{\rm c})\approx-\varepsilon/400$ changes $T_{\rm B}(N)$ for medium-sized and long chains by more than $5\%$.

According to renormalization-group calculations \cite{Duplantier_JPF} for long chains one should find $\Theta - T_{\rm B}\propto N^{-1/2}(\ln N/N_0)^{-7/11}$, where we would expect $N_0\approx 1$. Therefore, we fitted the Ansatz
\begin{equation}
q_l(N)=\Theta_l + \frac{a_1}{N^{1/2}(\ln N/N_0)^{7/11}} + \frac{a_2}{N^{2/2}} + \frac{a_3}{N^{3/2}} + \frac{a_4}{N^{4/2}}
\label{eq:tp4}
\end{equation}
and found even very good agreement for all 5 data sets. The nature of (\ref{eq:tp4}) dictates that $N_0$ has to be smaller than the smallest value of $N$ in the data set, but beyond this condition its precise value seems of minor importance.  For $N\ge2$ we obtain very good fits for a wide range of values with $N_0\lessapprox 0.8$ and also the values of $\Theta_l$ exhibit only minor variation of an order of $10^{-3}$. If data for small $N$ is excluded from the fit, larger values for $N_0$ lead to good fits as well.

It is not clear \emph{a priori} whether logarithmic corrections are actually traceable for the model we investigate with the system sizes considered here. To test this we also fitted the bare polynomial
\begin{equation}
q_p(N)=\Theta_p + \frac{b_1}{N^{1/2}} + \frac{b_2}{N^{2/2}} + \frac{b_3}{N^{3/2}} + \frac{b_4}{N^{4/2}},
\label{eq:p4}
\end{equation}
which also modeled the data very well albeit not as good as (\ref{eq:tp4}). Irrespective of the value of $r_{\rm c}$ we find $\Theta_p<\Theta_l$.

Finally we also fitted a polynomial of forth order to the inverse of the Boyle temperature assuming the relation
\begin{equation}
q_i(N)=\left(\frac1{\Theta_i} + \frac{c_1}{N^{1/2}} + \frac{c_2}{N^{2/2}} + \frac{c_3}{N^{3/2}} + \frac{c_4}{N^{4/2}}\right)^{-1},
\label{eq:hp4}
\end{equation}
which has to a lesser degree also been employed in the past \cite{Vogel}. Values for $\Theta_i$ are slightly larger but very similar to $\Theta_p$.

We compare all estimates for the $\Theta$-temperatures in Table \ref{tab:fit_Ttheta}. Although the choice of the higher-order correction terms used in eq.\ (\ref{eq:tp4}) can certainly be debated, at the very least we get an impression on how weakly the estimate of the $\Theta$-temperature for our data is influenced by the assumed analytical function. 

\begin{figure}
\begin{center}
\includegraphics[width=0.95\columnwidth]{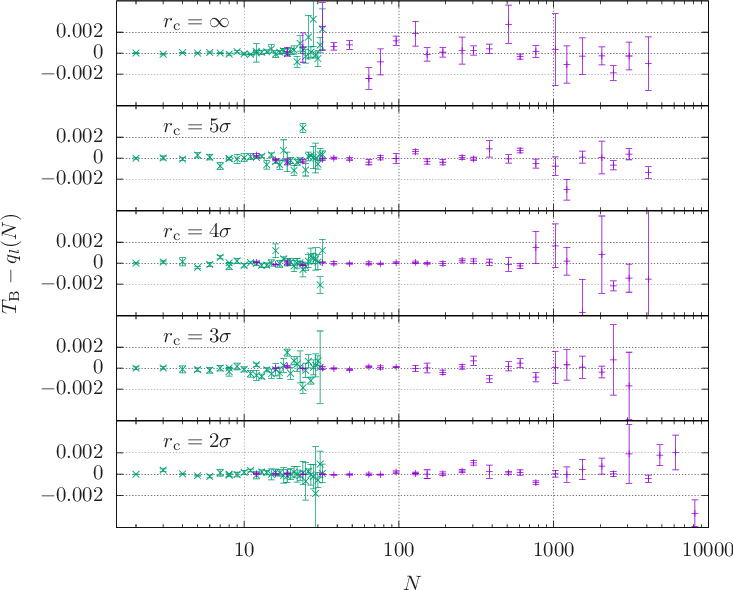}
\caption{\small{\label{fig:boyle_log_dev} Difference between measured Boyle temperatures and fitted functions $q_l$. Values are obtained by simple sampling ($\times$) and importance sampling ($+$).}}
\end{center}
\end{figure}

\begin{figure}
\begin{center}
\includegraphics[width=0.95\columnwidth]{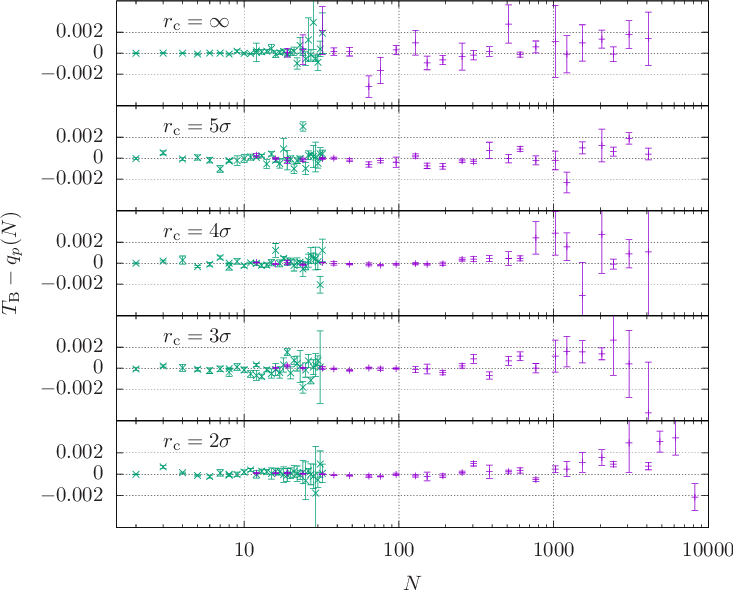}
\caption{\small{\label{fig:boyle_dev} Same as Fig.~\ref{fig:boyle_log_dev} for $q_p$.}}
\end{center}
\end{figure}

\begin{figure}
\begin{center}
\includegraphics[width=0.95\columnwidth]{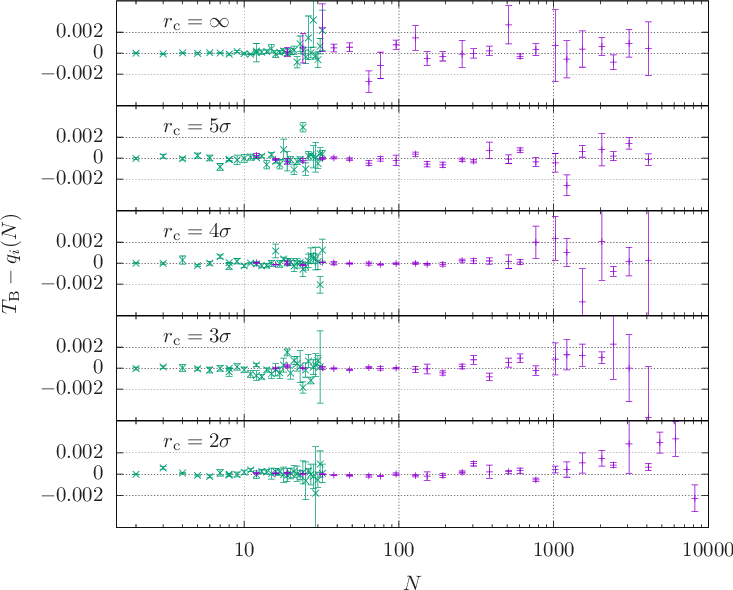}
\caption{\small{\label{fig:boyle_inv_dev} Same as Fig.~\ref{fig:boyle_log_dev} for $q_i$.}}
\end{center}
\end{figure}

\begin{figure}
\begin{center}
\includegraphics[width=0.95\columnwidth]{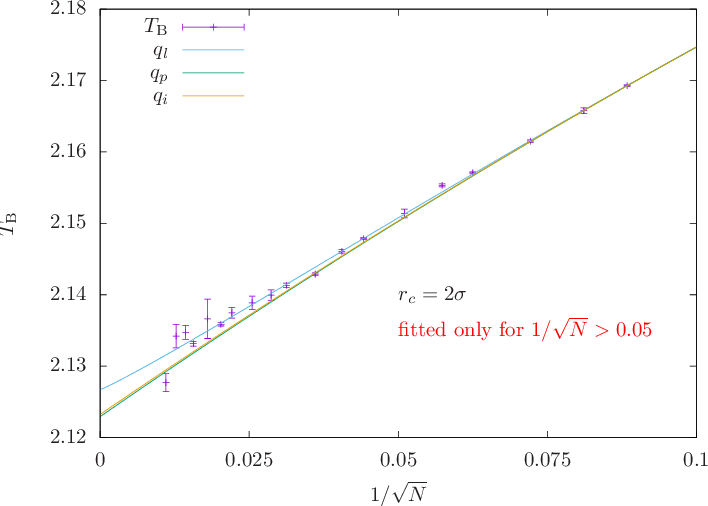}
\caption{\small{\label{fig:T_c2s} If the functions $q_l, q_p,$ and $q_i$ are fitted using only data for $N<400$ the Ansatz $q_l$ which includes logarithmic corrections best predicts the behavior of long chains while $q_p$ and $q_i$ (almost identical curves) generate too small values for $T_{\rm B}$.}}
\end{center}
\end{figure}

All three functions (\ref{eq:tp4}, \ref{eq:p4}, \ref{eq:hp4}) model the data very well and would be indistinguishable in a plot like Fig.~\ref{fig:boyle}. In order to ``zoom in'' and to better appreciate the (dis-)agreement the differences between the data and the fits are shown in Figs.~\ref{fig:boyle_log_dev}, \ref{fig:boyle_dev}, \ref{fig:boyle_inv_dev}. One can attempt to judge whether the data thus displayed show merely statistical noise by  randomly fluctuating around the zero value or whether they constitute a trend that indicates systematic deviations. It seems that fits of $q_p$ and $q_i$ both underestimate the Boyle temperatures for long chains and that $q_l$ represents the behavior best. For $r_{\rm c}=2\sigma$ the longest chains could be simulated and we fit in Fig.~\ref{fig:T_c2s} the three functional forms to the subset of smaller chains $N<400$ and used them as predictive models for the longer ones. While $q_p$ and $q_i$ now clearly underestimate the Boyle temperature, $q_l$ with the logarithmic corrections performs much better. Our results thus provide evidence for non-polynomial scaling; specific details, e.g., the exponent $7/11$ can, unfortunately, not be verified at this point.

\begin{table}
\caption{\small{\label{tab:fit_Ttheta} $\Theta$-temperatures obtained through fits of eqs. (\ref{eq:p4}), (\ref{eq:hp4}), and (\ref{eq:tp4}).}}
\begin{tabular}{|c|ccc|}  \hline
$r_{\rm c}$ & $\Theta_p$ & $\Theta_i$ & $\Theta_l$ \\ \hline
$\infty$  & $3.8921(3)$ & $3.8938(3)$ & $3.8979(5)$ \\
$5\sigma$ & $3.6907(4)$ & $3.6917(3)$ & $3.6954(9)$ \\
$4\sigma$ & $3.5368(2)$ & $3.5380(2)$ & $3.5427(3)$ \\
$3\sigma$ & $3.1761(3)$ & $3.1768(3)$ & $3.1814(4)$ \\
$2\sigma$ & $2.1236(2)$ & $2.1238(2)$ & $2.1265(2)$ \\ \hline
\end{tabular}
\end{table}

\begin{figure}
\begin{center}
\includegraphics[width=0.95\columnwidth]{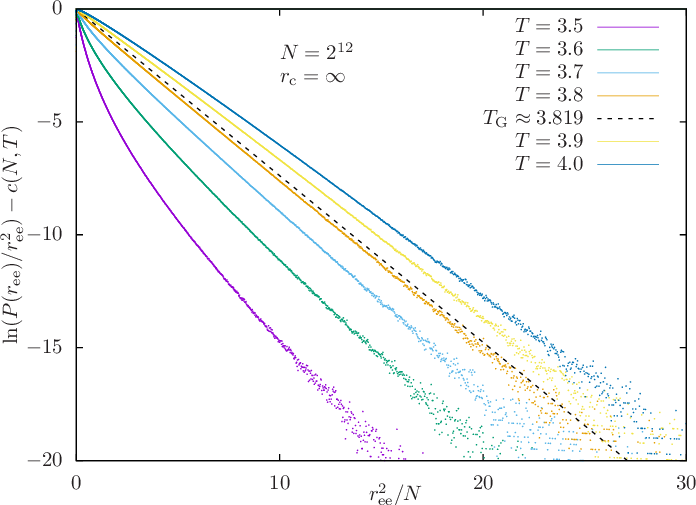}
\caption{\small{\label{fig:ree_distr} Fitting quadratic polynomials and interpolating the fitted coefficients of the linear and quadratic terms as functions of $T$ allows to estimate the temperature $T_{\rm G}$ where the relationship (\ref{eq:ree_distr}) is fulfilled. There, the second-order term vanishes and the remaining linear term leads to the dotted line. The polynomial's constant term $\text{c}(N,T)$ has been subtracted to allow easier visual comparison.}}
\end{center}
\end{figure}

Since the $\Theta$-point can also be determined by geometric means the value obtained through extrapolation of the Boyle temperatures can be tested. The method we use is similar to the technique employed in \cite{Yong}. At the transition the polymer behaves at large enough scales like a Gaussian random walk which implies that the end-to-end vector should be distributed according to
\begin{equation}
    P(\mathbf{r}_{\rm ee})\propto e^{-\mathbf{r}_{\rm ee}^2/{2\sigma_{r_{\rm ee}}(N)^2}}.
\end{equation}
This means that for the distribution of the end-to-end distance $r_{\rm ee}=|\mathbf r_{\rm ee}|$ we find
\begin{equation}
    \ln\left(P(r_{\rm ee})/r_{\rm ee}^2\right) = -r_{\rm ee}^2/{2\sigma_{r_{\rm ee}}(N)^2} + \text{c}.
    \label{eq:ree_distr}
\end{equation}
We, therefore, can define a ``Gau\ss'' temperature $T_{\rm G}$ as the temperature where the second-order term of a quadratic polynomial fitted to $\ln\left(P(r_{\rm ee})/r_{\rm ee}^2\right)$ as a function of $r_{\rm ee}^2$ vanishes (see Fig.~\ref{fig:ree_distr}).

\begin{figure}
\begin{center}
\includegraphics[width=0.95\columnwidth]{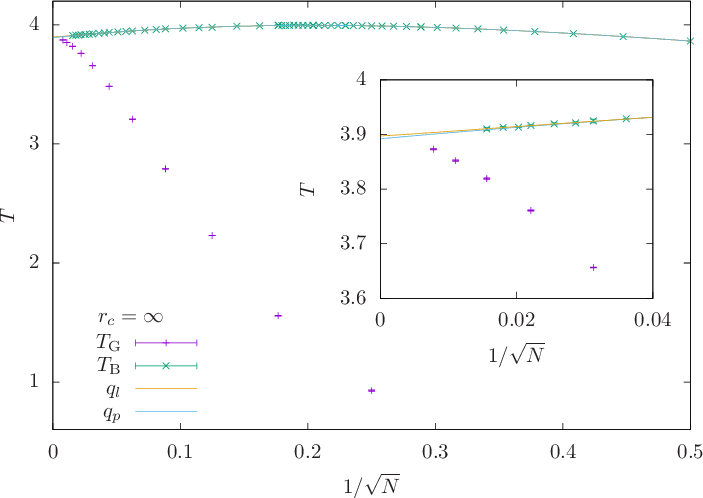}
\caption{\small{\label{fig:T_gauss} Comparing $T_{\rm G}$ to the Boyle temperatures shows good agreement in the thermodynamic limit. The inset magnifies details of the main plot for $N>600$.}}
\end{center}
\end{figure}

We measured $T_{\rm G}$ for $r_{\rm c}=\infty$ and various chain lengths and show the results together with the Boyle temperature in Fig.~\ref{fig:T_gauss}. For finite $N$ we find $T_{\rm G}<T_{\rm B}$. It is apparent that in the thermodynamic limit the different temperatures converge or become at least very close, but also that for any chain length $T_{\rm G}$ differs from $\Theta$ much more strongly than $T_{\rm B}$ does. Considering further that we do not know which scaling law $T_{\rm G}$ is supposed to follow we have to conclude that this geometric method of determining $\Theta$ is on its own less suitable than the second-virial-coefficient technique even though larger systems can be treated.

\subsection{The polymer at the $\Theta$-temperature}

For the investigation of the polymer in the tricritical region we use the temperature obtained with logarithmic corrections taken into account: $\Theta\equiv\Theta_l$.

\begin{figure}
\begin{center}
\includegraphics[width=0.95\columnwidth]{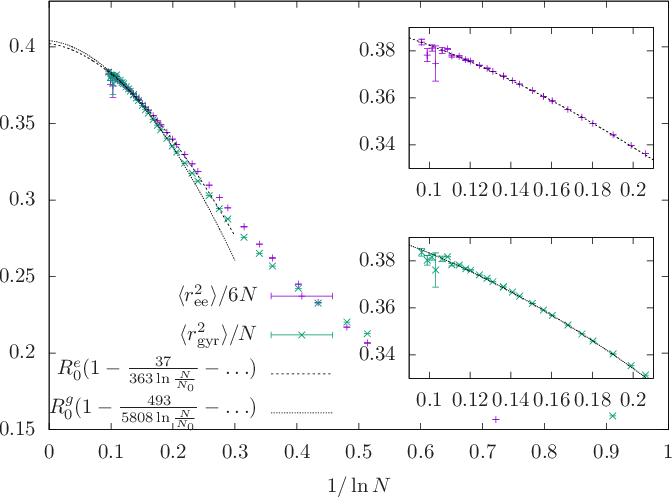}
\caption{\small{\label{fig:r_at_theta} Squared radius of gyration and end-to-end distance at the $\Theta$-temperature ($r_{\rm c}=\infty$). The insets show for $N>100$ the same data sets individually with their respective fits.}}
\end{center}
\end{figure}

\begin{figure}
\begin{center}
\includegraphics[width=0.95\columnwidth]{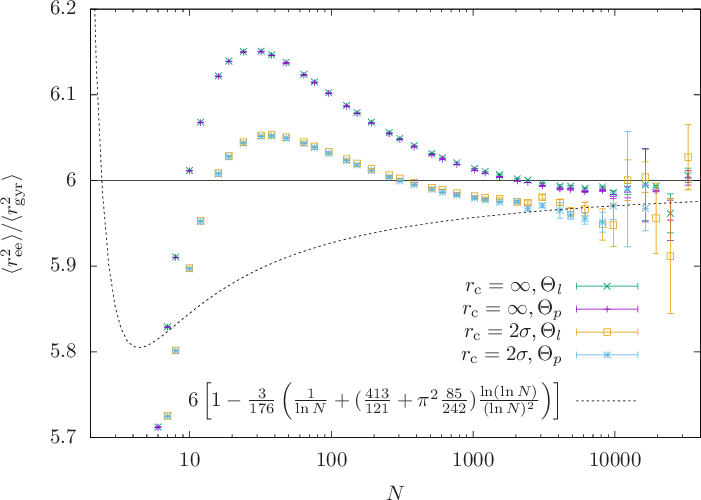}
\caption{\small{\label{fig:ratio} Ratio of squared end-to-end distance and squared radius of gyration at the $\Theta$-temperature and the analytical prediction.}}
\end{center}
\end{figure}

We performed simulations close to $\Theta$ for $r_{\rm c}=\infty$ and $r_{\rm c}=2\sigma$ to investigate scaling behavior. Here, chains with up to $N=32768$ are considered. It is well known and has been verified numerous times that the size of the chain measured by $\sqrt{\langle r_{\rm ee}^2\rangle}$ or $\sqrt{\langle r_{\rm gyr}^2\rangle}$ asymptotically scales like $\sqrt{N}$ as is expected from a Gaussian random walk. This means that for $\langle r_{\rm X}^2\rangle/N$ a non-zero limit for $N\rightarrow\infty$ exists. Unfortunately, logarithmic corrections which are expected here are generally difficult to verify and for the case of the tricritical $\Theta$-transition in particular Hager and Sch\"afer \cite{Hager} have shown that at least first-order ($\propto 1/\ln N$) and second-order ($\propto \ln(\ln N)/(\ln N)^2$) corrections are of the same magnitude for the chain length considered here and both have to be included in the analysis. They found that
\begin{eqnarray}
\frac{\langle r_{\rm ee}^2 \rangle}N = 6R_0\left(1 - \frac{37}{363\ln \frac{N}{N_0}}-\frac{37\kappa\ln(\ln \frac{N}{N_0})}{363\ln^2\frac{N}{N_0}} \right. \nonumber \\
                                      + O\left((\ln N)^{-2}\right) \Bigg), \label{eq:re} \\
\frac{\langle r_{\rm gyr}^2 \rangle}N = R_0\left(1 - \frac{493}{5808\ln \frac{N}{N_0}}-\frac{493\kappa\ln(\ln \frac{N}{N_0})}{5808\ln^2\frac{N}{N_0}} \right. \nonumber \\
                                      + O\left((\ln N)^{-2}\right) \Bigg), \label{eq:rg}
\end{eqnarray}
and consequently
\begin{eqnarray}
\frac{\langle r_{\rm ee}^2 \rangle}{\langle r_{\rm gyr}^2 \rangle} = 6\left(1 - \frac{3}{176\ln \frac{N}{N_0}}-\frac{3\kappa\ln(\ln \frac{N}{N_0})}{176\ln^2\frac{N}{N_0}} \right. \nonumber \\
                                      + O\left( (\ln N)^{-2}\right) \Bigg), \label{eq:ratio}
\end{eqnarray}
where $\kappa=\frac{413}{212}+\pi^2\frac{85}{242}\approx 6.88$ is a numerical constant. In Fig.~\ref{fig:r_at_theta} we attempt to fit these relations to $\langle r_{\rm ee}^2 \rangle/6N$ and $\langle r_{\rm gyr}^2 \rangle/N$ for $(\ln N)^{-1}<0.2$, i.e., for $N>148$. We included a term $\propto(\ln N)^{-2}$.  If we set $N_0\approx1$ the individual fits achieve rather good agreement with the data albeit not within error bars. Besides, the respective estimates for $R_0$ differ by about $0.5\%$: $R_0^e\approx0.402$ vs $R_0^g\approx0.404$.
Although our data is comparatively precise we are still unable to verify or reject the predictions. We can at least note that our results are not in conflict with theory and we find no reason to doubt that data for larger chains would follow eqs. (\ref{eq:re}, \ref{eq:rg}) more closely converging to the same limit $R_0$.
In Fig.~\ref{fig:ratio} we show the ratio $\langle r_{\rm ee}^2\rangle/\langle r_{\rm gyr}^2\rangle$ together with the prediction (\ref{eq:ratio}) as a function of $\ln N$. For the latter the argument should actually be $N/N_0$ instead of $N$ with an unknown constant $N_0$ leading to a horizontal shift of the curve. Regardless of this issue, the statistical errors are too large and the sizes still too small to draw definite conclusion. It is clear that asymptotic behavior can only be observed for long chains and our data shows that for this model precise data with at least $N>10000$ is needed. We show data for both $\Theta_l$ and $\Theta_p$ to illustrate that the uncertainty about the true value of $\Theta$ plays a very minor role when it comes to the agreement or disagreement with theoretical predictions.

\begin{figure}
\begin{center}
\includegraphics[width=0.95\columnwidth]{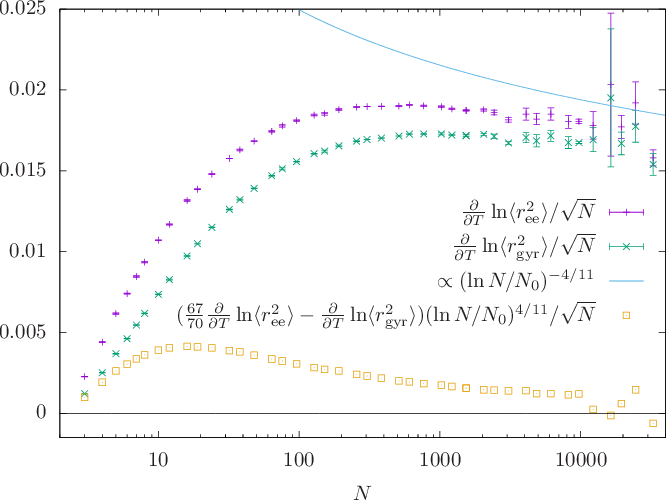}
\caption{\small{\label{fig:r2_deriv_at_theta} Temperature derivative of the logarithmic squared radius of gyration and end-to-end distance at the $\Theta$-temperature and the predicted asymptotic scaling ($r_{\rm c}=\infty$).}}
\end{center}
\end{figure}

For the temperature derivative of the squared radius of gyration and end-to-end distance the calculations of Duplantier \cite{Duplantier_JCP} predict
 \begin{equation}
 \frac{\partial}{\partial T} \ln \langle r_X^2 \rangle \sim \frac{\sqrt N}{(\ln N/N_0)^{4/11}}
 \end{equation}
and
 \begin{equation}
 \frac{ \frac{\partial}{\partial T} \ln \langle r_{\rm ee}^2 \rangle }{ \frac{\partial}{\partial T} \ln \langle r_{\rm gyr}^2 \rangle }= \frac{70}{67}
 \end{equation}
for long chains. We took derivatives of quadratic polynomials fitted to the canonical averages at temperatures close to the $\Theta$-temperature: $|\Theta-T|\le0.02$. Using linear polynomials instead produced almost identical results which suggests that this procedure is valid. In Fig.~\ref{fig:r2_deriv_at_theta} we display the derivatives of the logarithmic averages along with the rescaled difference 
\begin{equation}
\left(\frac{67}{70}\frac{\partial }{\partial T}\ln \langle r_{\rm ee}^2 \rangle - \frac{\partial }{\partial T}\ln \langle r_{\rm gyr}^2 \rangle \right)\left(\ln \frac N{N_0}\right)^{4/11}/\sqrt{N}
\label{r2_deriv_diff}
\end{equation}
setting $N_0=1$. For the latter we are unable to estimate statistical uncertainties since the data for $\langle r_{\rm ee}^2 \rangle$ and $\langle r_{\rm gyr}^2 \rangle$ is highly correlated. We find that for long enough chains $\frac{\partial}{\partial T} \ln \langle r_X^2 \rangle/\sqrt{N}$ varies only very little, even though a decrease for $N>1000$ can be observed and the data might approach the predicted asymptotic behavior. However,  there is too much noise and the chains are still to short to be certain. It also seems likely that the difference (\ref{r2_deriv_diff}) approaches zero as predicted, albeit very slowly. On the one hand this observation supports the theory, while on the other it demonstrates that even the behavior of the chains we investigate here which are of substantial length is still considerably different from that of the infinite chain.

%

\begin{figure}
\begin{center}
\includegraphics[width=0.95\columnwidth]{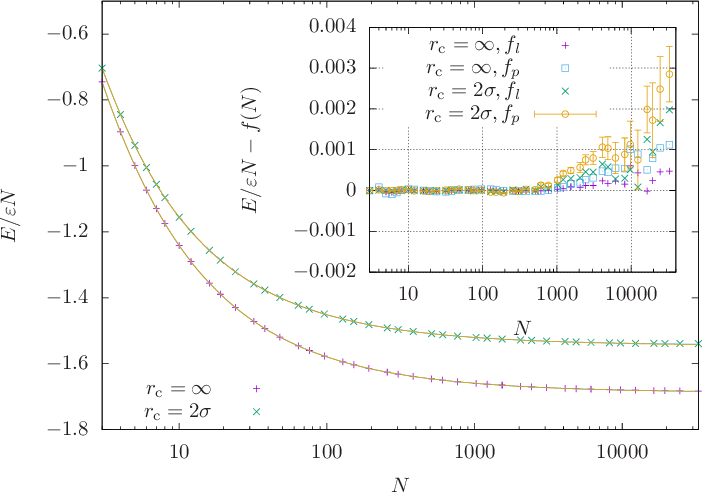}
\caption{\small{\label{fig:E_at_theta} Energy per monomer as function of system size at the $\Theta$-temperature for $r_{\rm c}=\infty$ and $r_{\rm c}=2\sigma$. The inset shows the deviation of the data from the fitted function $f_l$ and $f_p$ from eqs.\ (\ref{E_fit_b}) and (\ref{E_fit_a}), respectively. Error-bars are displayed exemplarily only for one data set in order to preserve visibility of all data points.}}
\end{center}
\end{figure}

How the energy per monomer depends on $N$ at $\Theta_l$ is plotted in Fig.~\ref{fig:E_at_theta}. We fitted the two functions
\begin{eqnarray}
f_l(N) &=& e_l + \frac{\beta_1}{N^{1/2}\left(\ln(N/N_0)\right)^{4/11}} + \frac{\beta_2}{N^{2/2}} + \dots,
\label{E_fit_b}\\
f_p(N) &=& e_p + \frac{\alpha_1}{N^{1/2}} + \frac{\alpha_2}{N^{2/2}} + \dots
\label{E_fit_a}
\end{eqnarray}
to the data with the predicted asymptotic behavior given by the first two terms of $f_l$ \cite{Cloizeaux}.
Again, the value of $N_0$ is restricted by the smallest $N$ in the data and it does not change the quality of the fit by much.
In any case, the functional form with logarithmic corrections fits the data better than the bare polynomial.
For Fig.~\ref{fig:E_at_theta} all data are included ($N\ge3$) and $N_0=0.3$.
With these conditions and $r_{\rm c}=\infty$ we obtain $e_l=-1.68774(4)$ and $e_p=-1.6890(1)$ while for $r_{\rm c}=2\sigma$ it is $e_l=-1.5445(1)$ and $e_p=-1.5460(2)$.

\begin{figure}
\begin{center}
\includegraphics[width=0.95\columnwidth]{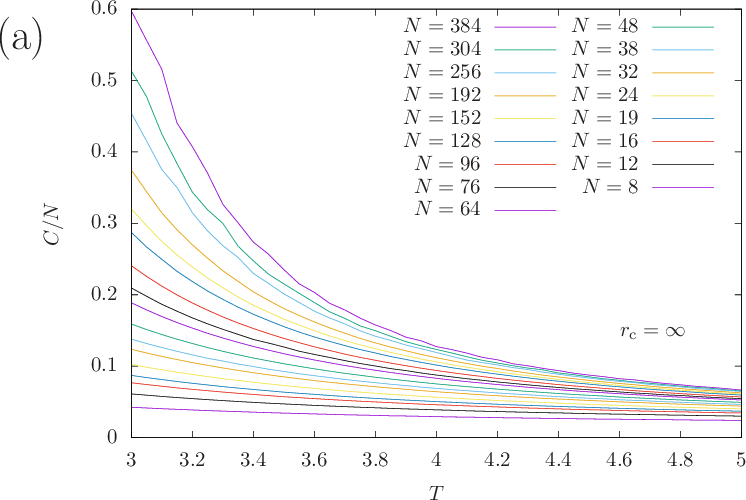}
\includegraphics[width=0.95\columnwidth]{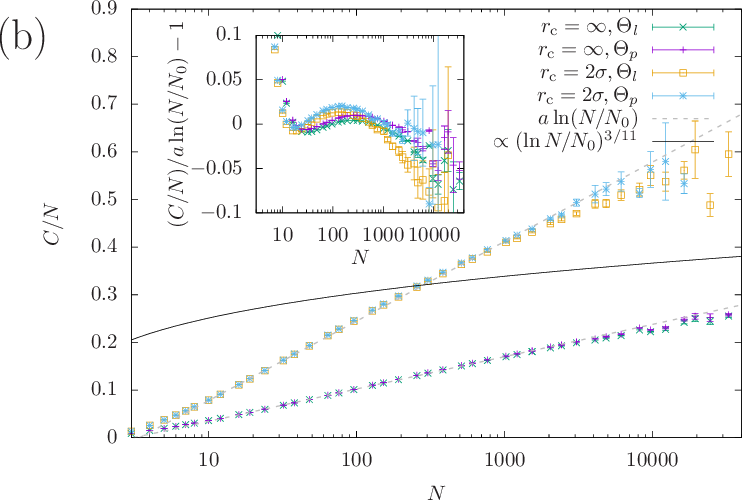}
\caption{\small{\label{fig:C_at_theta} (a) Specific heat per monomer $C/N$ as function of temperature $T$ for different sizes $N$. (b) Specific heat per monomer $C/N$ at the $\Theta$-temperature for different sizes $N$. As a visual reference we included the predicted asymptotic scaling using $N_0=1$ (black line). Although the first impression is that $C/N$ scales linearly with $\ln(N/N_0)$ (dashed lines), a closer inspection (inset) reveals systematic deviations which increase with system size. Here, $a$ is a fitting parameter. The symbols in the inset belong to the same parameter combinations as in the main plot.}}
\end{center}
\end{figure}

We used the opportunity to successfully verify our results for the specific heat as function of temperature shown in Fig.~\ref{fig:C_at_theta}a  by comparing to an earlier study \cite{Baumgaertner} of the same model \footnote{In this study the definition of the second virial coefficient for a gas of monomers was used which led to Boyle temperatures different from the values presented here.}. At the $\Theta$-temperature the specific heat is predicted \cite{Duplantier_EPL} to behave as $C/N\propto(\ln N/N_0)^{3/11}$. At first sight our data in Fig.~\ref{fig:C_at_theta}b shows a very different picture and a linear dependence on $\ln N$, i.e., Gaussian chain behavior, seems to be apparent.
However, deviations from this behavior are present at all sizes (see inset) and the scaling finally changes at $N\approx10^4$ to what might very well be the predicted behavior.
We also see that small changes in the temperature, e.g., using $\Theta_p$ instead of $\Theta_l$ hardly changes the scaling.

\begin{figure}
\begin{center}
\includegraphics[width=0.95\columnwidth]{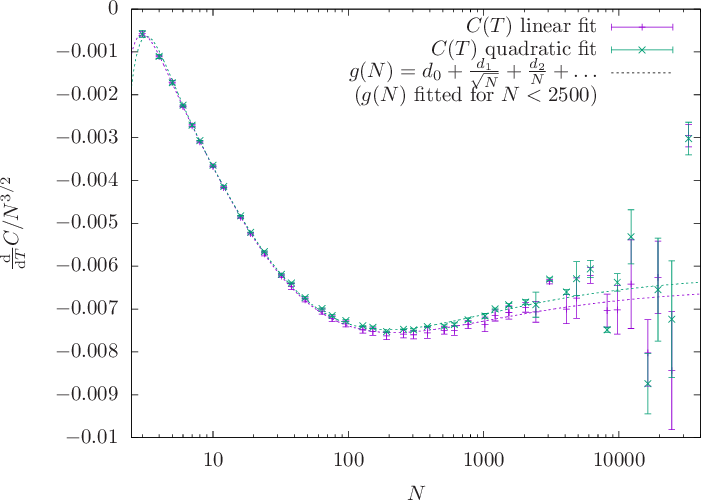}
\caption{\small{\label{fig:C_deriv_at_theta} Temperature derivative of the specific heat $\diff C/\diff T$ at the $\Theta$-temperature for different chain lengths $N$. }}
\end{center}
\end{figure}

Finally we measured the temperature derivative of the specific heat at $\Theta$. This was done by fitting polynomials of first and second order to $C(T)$ on an interval $[3.85,3.93]$ containing $\Theta$ and taking their derivatives. Because for each size $N$ only few data points (temperatures) are available the error estimates themselves which were obtained through the fitting process are relatively uncertain. The results are shown in Fig.~\ref{fig:C_deriv_at_theta}. To our knowledge there is no analytical prediction for $\diff C(T)/\diff T$ since the partition function has not been developed beyond second order. The data suggest that in the limit of long chains $\diff C(T)/\diff T|_{T=\Theta}/N\propto N^{1/2}$, however, a straightforward fit of a polynomial in powers of $N^{-1/2}$ to $\diff C(T)/\diff T|_{T=\Theta}/N^{3/2}$ does not work very well and we suspect that logarithmic corrections could play a role here as well.

\section{Conclusion}
\label{sec-conc}

In this study we applied the algorithm we introduced in a recent publication \cite{poly_tree} to investigate a polymer model at the $\Theta$-transition. Within the framework of the algorithm decisions on proposed updates are made with as little information about the associated change in energy as possible. Since the precise evaluation of the change in energy which is computationally very expensive is thus avoided, the method is very fast. It is now possible to simulate chains in continuous space with ten thousands of monomers even if the potential is not truncated and each monomer interacts with all others. The accessible chain lengths are now more than an order of magnitude larger than what has been achieved previously in similar studies \cite{Rubio1} and are comparable to what has been done for lattice models with short-range next-neighbor interaction \cite{GrassbergerHegger,Grassberger}. Although the model we have used here is rather simple, the algorithm is versatile and different pair potentials, elastic bonds, stiffness, or heterogenity can easily be incorporated.

True to the philosophy of the algorithm we developed a technique to accurately and efficiently measure expectation values of geometric and thermodynamic observables which allows for uncertainties in the majority of individual measurements. Although the underlying idea is rather simple and it seems likely that it has been found before, we are unaware of any previous application.

We measured the second virial coefficient in order to determine the Boyle temperature and systematically investigated how the latter is influenced by a truncation of the monomer-monomer interaction potential.
With the exception of a very small cutoff distance, with increasing size the Boyle temperature increases, reaches a maximum, and approaches the infinite chain limit, i.e., the $\Theta$-temperature from above.
Truncation of the potential has a big influence and reduces the Boyle temperature up to about 50\%.
Data for all cutoff radii can be very well modeled with polynomials of fourth order, yet when the linear term is modified with the predicted logarithmic correction the agreement is better still.

At $\Theta$-temperature the scaling of the average end-to-end distance and radius of gyration is not in conflict with Duplantier's predictions, but the agreement is also not as good as one might wish. It seems that the chain lengths we were able to simulate are still too small. On the other hand we do observe that the relationship between the temperature derivatives of the two quantities compares very well with theory. The specific heat again does not clearly demonstrate the predicted logarithmic asymptotic scaling. However, we were able to reach system sizes where the linear scaling that we observe for short chains brakes down and the curve bends in the right direction.
Unfortunately, considering that several hundreds of CPUs ran for several months to produce the data presented here, we are not overly optimistic about the possibility of investigating much longer chains unless additional substantial improvements can be made to the methodology.

\section*{Acknowledgements}
The project was funded by the Deutsche Forschungsgemeinschaft (DFG, German Research Foundation) through the Collaborative Research Centre under Grant No. 189\,853\,844--SFB/TRR 102 (project B04).

\bibliography{LJ_theta}

\end{document}